\documentclass[aps, prd, superscriptaddress, reprint, amsmath, amssymb, showkeys, floatfix, nofootinbib,noeprint]{revtex4-2}
\usepackage{graphicx}% Include figure files
\usepackage{dcolumn}% Align table columns on decimal point
\usepackage{bm}% bold math
\usepackage{natbib}
\usepackage{makecell}
\usepackage[colorlinks,linkcolor=blue,urlcolor=blue,citecolor=blue]{hyperref}
\usepackage{booktabs}
\usepackage{siunitx}
\usepackage{rotating}
\usepackage{multirow}
\usepackage{verbatim}
\usepackage{comment}
\usepackage{color}

\begin{document}

\title{Neutron Star Merger Universality Relations for a Quark--Hadron Crossover \\Equation of State}

\author{Wei Sun}
\email{wsun3@nd.edu}
\affiliation{Center for Astrophysics, Department of Physics and Astronomy, University of Notre Dame, \\Notre Dame, IN 46556, USA}
\author{Grant J. Mathews}
\email{gmathews@nd.edu}
\affiliation{Center for Astrophysics, Department of Physics and Astronomy, University of Notre Dame, \\Notre Dame, IN 46556, USA}
\author{Atul Kedia}
\email{akedia@nccu.edu}
\affiliation{Department of Mathematics and Physics, North Carolina Central University, Durham, NC 27707, USA}
\date{\today}

\begin{abstract}
A crossover transition from hadronic to quark matter during the merger of neutron stars can lead to interesting observational consequences in the emergent gravitational radiation. In particular, the increased pressure in the crossover density region ($2-5$ times the nuclear saturation density) can lead to an extended duration of high-frequency ($\sim 2-3.5$ kHz) gravitational wave emission during the post-merger epoch. We study the universality relations for a variety of equations of state with and without a crossover transition to a quark mater based upon the QHC19 and QHC21 EoSs.
%updated QHC21 EoS which has been developed on the basis of NICER observations suggesting the possibility of larger radii for the most massive neutron stars. 
We find that the long-duration post-merger gravitational-wave emission is more pronounced in the QHC21 EoS. We then consider universality relations among various observable quantities that have been identified in simulations with hadronic equations of state.  We show that there are deviations from the universality relations for some quantities that may ultimately provide an observational signal that such a transition has occurred.
\end{abstract}


\maketitle

%*********************************************************************
%******************** INTRODUCTION *************************************
%***************                                     *****************
\section{Introduction}
\label{introduction}

Neutron stars have an interior energy density exceeding the nuclear saturation density. As such, neutron stars may undergo a deconfinement transition from hadronic nuclear matter to quark matter \cite{Annala2020, Christoph2026, Gao2024, Bastian2021, Hammond2026, Raithel2023, Shamim2024, Chatterjee2026}. Indeed, this transition has been inferred in ground-based heavy-ion collision experiments \cite{Chiavassa20}. However, a transition to quark matter has not yet been observed in the context of astrophysical observations. 

The absence of accurate first-principle predictions at densities beyond the nuclear saturation density has thus far hindered the determination of the phase of matter inside neutron star (NS) cores. The detection of the binary neutron star (BNS) merger event, GW170817 \cite{LIGO2017, LIGO2019}, by LIGO, motivated the question as to whether evidence for the formation of quark matter could be obtained from gravitational wave (GW) detectors \cite{Prakash2024, Huang2022, Atul2022, Hammond2026, Raithel2023, Shamim2024, Espino2024, Chatterjee2026, Bauswein2012, Bauswein2015, Bauswein2019, Bauswein2020, Bauswein2025}.

Neutron star matter remains in the hadronic phase up to densities in excess of $n_{\rm sat}$, where $n_{\rm sat} \simeq 0.16~\mathrm{fm}^{-3}$ is the nuclear saturation density. In this regime, modern nuclear theoretical tools, such as chiral effective field theory ($\chi$EFT)\cite{ChiralEFT2013, ChiralEFT2020, Hu2026} provide a reasonable description of the equation of state (EoS). Similarly, at extremely high densities $\gtrsim 40~ n_{\rm sat}$, perturbative-QCD (pQCD) \cite{pQCD2018, Kurkela2010} offers a reasonable approximation to the quark-matter EoS.

The behavior of the squared speed of sound $c_s^2$ provides valuable insight into the EoS of NS matter. The conformal limit corresponds to \(c_s^2 = 1/3\), and high-density quark matter is expected to approach this asymptotically. In contrast, $\chi$EFT predicts \(c_s^2 < 1/3\) below the nuclear saturation density. However, at intermediate densities relevant to NS cores, first-principle calculations are not available. This has motivated the development of model-independent interpolation schemes and statistical inference methods to construct various versions of viable EoSs of NS-matter \cite{NSConsMost, NSConsAnnala, Annala2020}. These EoSs should encompass the widest possible range of physically realistic EoSs, including different types of phase transitions (PT) and possible onset densities of quark matter. Bayesian analyses incorporating current astrophysical observations indicate a high posterior probability that massive neutron stars undergo a transition to quark matter \cite{Annala2023,EoS_inference2024}. However, whether this transition is first-order or proceeds as a smooth crossover remains an open question.

It is possible that neutron stars undergo a strong first-order phase transition (FOPT) \cite{Gao2024, Bastian2021}. This could lead to a reduction in pressure support during a coexistence phase as hadronic matter is converted into quark matter. Consequently, the transition to a quark-matter core could accelerate the collapse of the remnant into a black hole (BH) \cite{Bauswein2019, Elias2019} (see however Ref.~\cite{Fujimoto2023} where a high pressure FOPT was constructed). In Ref.~\cite{Atul2022, Huang2022} it was noted that a crossover transition could lead to a long lived hypermassive neutron star (HMNS). Such remnants can be an important site for nucleosynthesis \cite{Sasaki24} and provide a plausible explanation for the optical emission observed in kilonovae \cite{Chatziioannou2025}. It also can lead to a long duration $\sim 2-3$ kHz gravitational wave emission \cite{Atul2022, 2021PhRvD.104h3029P, Mathews2026}.

Recent NICER observations \cite{Miller_2021,Riley_NICER_2021} suggest small variation in the neutron star radius as the mass increases from $1.4$ to $2.1\,M_{\odot}$. This implies a rapid increase in pressure with density in the range of $\sim 2$--$5\,n_{\rm sat}$. Motivated by these observations, the previous Quark--Hadron Crossover (QHC19) EoS \cite{Baym2019} was modified to the QHC21 \cite{Kojo2022} family of hybrid crossover models, such that the transition to quark matter occurs at a lower density. The crossover to quark matter stiffens the EoS at densities of $\sim 2$--$5\,n_{\rm sat}$, providing additional pressure support for the merger remnant and thereby extending the lifetime of the HMNS. This prolonged lifetime has been proposed as a distinguishing signature of crossover hybrid EoSs compared to soft purely hadronic EoSs and hybrid EoSs with a FOPT \cite{Atul2022, Huang2022}.

In addition to these developments, universality relations have been proposed across a broad range of hadronic EoSs \cite{Bauswein2012, Bauswein2015, Takami2015, Raithel2022, Breschi2024}. These universality relations establish correlations among quantities relevant to GW observations and the neutron star EoS. In this work, we consider the EoS dependence of the dominant post-merger GW frequency $f_{\rm peak}$ (also referred to as $f_2$ in the literature, Refs.~\cite{Bernuzzi2015, Takami2015, Huang2022, Hensh2025, Huang2026}), the GW frequency at maximum strain amplitude $f_{\rm max}$, and the dimensionless tidal deformability parameter $\Lambda$. We also consider the characteristic mean-density scale $(M_{\rm grav}/R_{\rm max}^{3})^{1/2}$ and the maximum of $\rho_{\rm max}(t)$ during the first $5$ milliseconds after merger, $\rho_{\rm max}^{\rm max}$, where $\rho_{\rm max}(t)$ denotes the maximum rest-mass density at time $t$.

More recently, deviations from the hadronic universality relations have been proposed as a means of identifying a FOPT in neutron star mergers \cite{Bauswein2019, Bauswein2020, Bauswein2025, Prakash2024}. These works utilized different parameterizations of the DD2F-SF EoS \cite{Bastian2021}, which contain a strong first-order PT. The pronounced softening of these EoSs leads to a discontinuous jump in the baryon density. As a result, the transient merger remnant collapses to a black hole on very short timescales. The pre-collapse remnant exhibits a higher value of $f_{\rm peak}$, leading to a clear deviation from the otherwise universality relation for $f_{\rm peak}$--$\Lambda$. The relation between $\rho_{\rm max}^{\rm max}$ and $f_{\rm peak}$ was also investigated. These works demonstrate that deviations from hadronic universality relations can provide signatures of a strong FOPT.

However, whether a crossover quark-hadron transition produces identifiable signatures in these universality relations has not yet been systematically explored. In the present work, we address this question by performing simulations using both the QHC19 and the updated QHC21 models. The QHC models smoothly interpolate between the low-density nucleonic Togashi \cite{Togashi2017} or $\chi$EFT \cite{Drischler2021, Drischler2021review} EoSs and the high-density Nambu--Jona-Lasinio (NJL) quark matter model \cite{Vogl1991, Klevansky1992}. Different QHC parameterizations correspond to different onset densities of quark matter and different quark matter coupling strengths, resulting in varying degrees of EoS stiffness. In addition, we perform a set of simulations with purely hadronic EoSs to establish the baseline hadronic universality relations.

Our work is complementary to two other recent works that also considered the effects of a QHC crossover EoS on universality. In particular, Ref.~\cite{Hensh2025} studied the relation between the frequency of the dominant post-merger peak ($\sim f_{peak}$ in our notation), and tidal deformability ($\Lambda$). Ref.~\cite{Huang2026} quantified how tightly can $f_{peak}$ be determined. Our work differs from these studies in that we consider a number of different universality relations. We also consider a wider range of stiffness in the hadronic EoSs to construct the universality relations.

The paper is organized as follows. Section~\ref{themodel_eos} discusses the microscopic EoSs used in the merger simulations. Section~\ref{themodel_hydrodynamics} describes the numerical setup of the simulations. Section~\ref{themodel_gw} presents the GW analysis, with particular emphasis on the spectral properties of the post-merger signal. The simulation results are presented in Section~\ref{results}, including the spectral properties in Section~\ref{results_psd} and, most importantly, the universality relations for the crossover EoSs in Section~\ref{results_universality}. Finally, Section~\ref{summary} summarizes our findings.

%*********************************************************************
%******************** INTRODUCTION *************************************
%***************                                     *****************

%=====================================================================
%=================== FIG. 1 M-R Relations of EoS =====================
% I put the Fig.1 (M-R relations of EoS) here because I want it to be shown at the top of Page2.
\begin{figure*}[!htbp]
\centering
\includegraphics[width=1\linewidth]{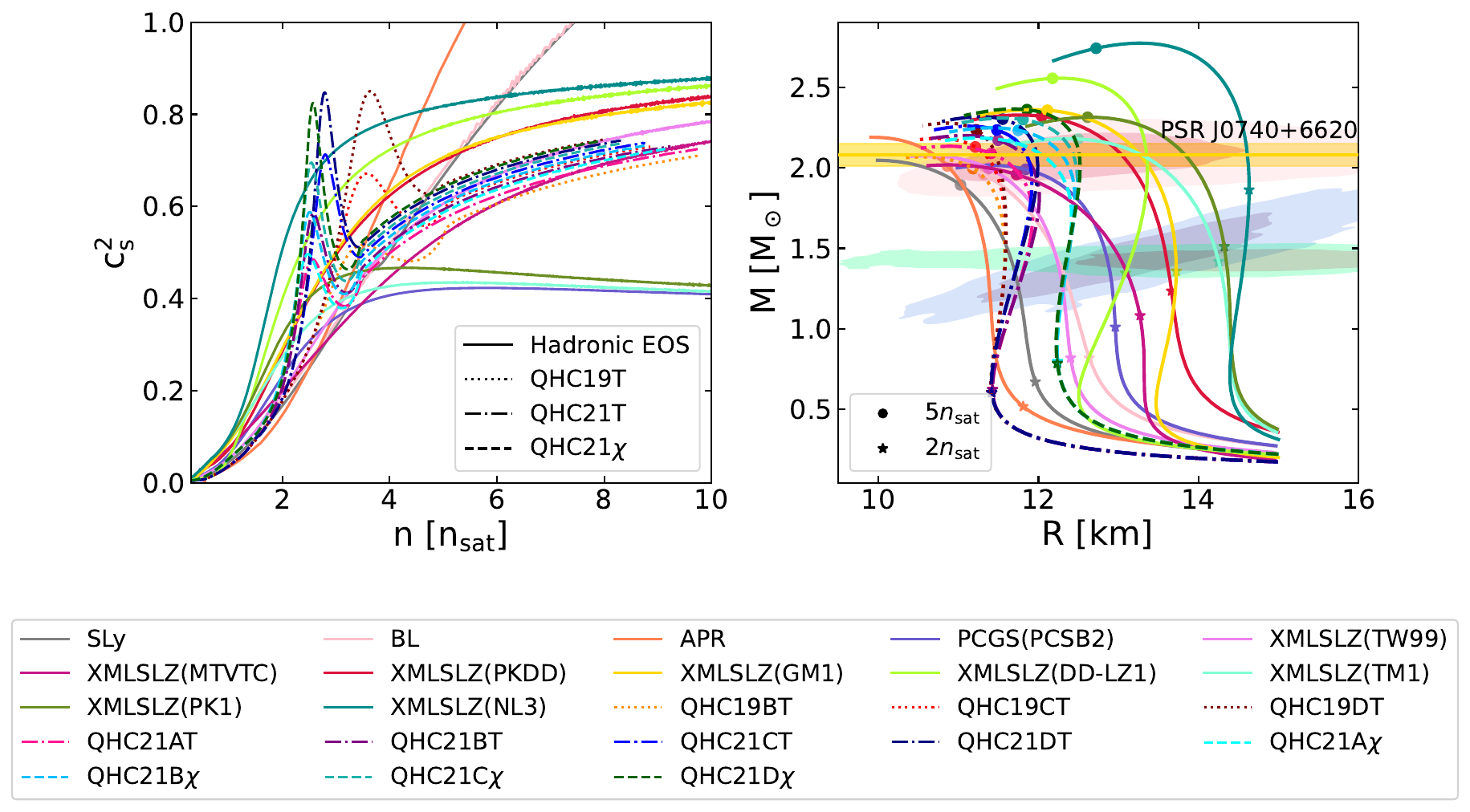}
\caption{\label{fig:mr}Comparison of the employed hadronic and hybrid EoSs. (Left) Squared speed of sound, $c_{s}^{2}$, as a function of baryon number density, $n$. (Right) Gravitational mass--radius relations for TOV stars. The star and circle points indicate central densities of $2n_{\rm sat}$ and $5n_{\rm sat}$, respectively. The NICER mass constraint for PSR J0740+6620, the most massive neutron star currently known with a precise mass measurement, is also shown \cite{Fonseca2021}. The NICER $M$--$R$ posterior distributions for PSR J0030+0451 (blue) \cite{Miller2019}, PSR J0740$+$6620 (pink) \cite{Dittmann2024}, and PSR J0437-4751 (green) \cite{Miller2026} are also shown, with $1\sigma$ and $2\sigma$ confidence regions.}
\end{figure*}
%===========================================================================
%***************************************************************************************
%===========================================================================
%=================== TABLE. 1 TOV Properties of EoS ========================
% I put the Tab.1 here because I want it to be shown at the top of Page3.
\begin{table*}[!htbp]
  \caption{Mass properties of the neutron stars, including the baryonic mass $M_{\rm baryon}$, the gravitational mass $M_{\rm grav}$ from the Tolman–Oppenheimer–Volkoff (TOV) solution, and the total ADM mass $M_{\rm ADM}$ of the binary system with an initial separation of 45~km. The values of $M_{\rm grav}$ and $M_{\rm ADM}$ are computed using the piecewise-polytropic representations of the EoSs. Also listed are the characteristic GW frequencies during the merger and post-merger phases: the dominant post-merger GW frequency $f_{\rm peak}$, extracted from the power spectral density, and the GW frequency at the time of maximum strain amplitude, $f_{\rm max}$. Masses are given in units of $M_\odot$, and frequencies are given in kHz.}
%Mg is for raw EOSs, TOV, ADM is PP in Lorene.
  \label{tab:tab_mass}
  \begin{ruledtabular}
  \begin{tabular}{@{}lccccc@{\qquad}|lccccc@{}}
    \makecell{Hybrid\\EOS}    & \(M_{\rm{baryon}}\) &\(M_{\rm{grav}}\) & \(M_{\rm{ADM}}\) & \(f_{\rm{peak}}\) & \(f_{\rm{max}}\) 
    & \makecell{Hadronic\\EOS} & \(M_{\rm{baryon}}\) &\(M_{\rm{grav}}\) & \(M_{\rm{ADM}}\) & \(f_{\rm{peak}}\) & \(f_{\rm{max}}\)\\
    \hline
    QHC21A$_\chi$\cite{Kojo2022} & 1.45   & 1.327 & 2.63  & 2.74 & 1.66 & SLy\cite{Douchin2001}           &1.45 &1.323 & 2.62 &3.29 &1.92\\
    QHC21A$_\chi$ & 1.50   & 1.364 & 2.71  & 2.81 & 1.83 & SLy           &1.50  &1.363 & 2.70 &-- &1.90 \\
    QHC21A$_\chi$ & 1.55   & 1.402 & 2.79  & 2.90 & 1.87 & SLy           &1.55 &1.402 & 2.78 &-- &1.88\\
    \hline
    QHC21B$_\chi$\cite{Kojo2022} & 1.45   & 1.328 & 2.63  & 2.71 & 1.67 & BL\cite{Bombaci2018, Carreau2019, BLzenodo}             &1.45 &1.325 & 2.63 &3.13 &1.85\\
    QHC21B$_\chi$ & 1.50   & 1.369 & 2.71  & 2.81 & 1.83 & BL             &1.50  &1.371 & 2.71 &3.42 &1.82\\
    QHC21B$_\chi$ & 1.55   & 1.410 & 2.79  & 2.84 & 1.86 & BL             &1.55 &1.407 & 2.79 &-- &1.90\\
    \hline
    QHC21C$_\chi$\cite{Kojo2022} & 1.45   & 1.323 & 2.63  & 2.71 & 1.68 & APR\cite{Akmal1998, Haensel2007, Davis2024, APRzenodo}            &1.45 &1.316 & 2.61 &3.29 &1.98\\
    QHC21C$_\chi$ & 1.50   & 1.368 & 2.71  & 2.74 & 1.86 & APR            &1.50  &1.353 & 2.69 &3.52 &2.03\\
    QHC21C$_\chi$ & 1.55   & 1.414 & 2.79  & 2.81 & 1.85 & APR            &1.55 &1.401 & 2.77 &3.77 &2.00\\
    \hline
    QHC21D$_\chi$\cite{Kojo2022} & 1.45   & 1.322 & 2.63  & 2.68 & 1.67 & PCGS(PCSB2)\cite{Hempel2010, Pradhan2022, PCGSzenodo}    &1.45 &1.333 & 2.64 & &\\
    QHC21D$_\chi$ & 1.50   & 1.371 & 2.71  & 2.74 & 1.83 & PCGS(PCSB2)    &1.50  &1.374 & 2.72 &2.94 &1.72\\
    QHC21D$_\chi$ & 1.55   & 1.420 & 2.79  & 2.78 & 1.88 & PCGS(PCSB2)    &1.55 &1.415 & 2.80 &-- &1.80\\ 
    \hline
    QHC21AT\cite{Kojo2022}       & 1.45   & 1.324 & 2.61  & 2.90 & 1.74 & XMLSLZ(TW99)\cite{xmlslz_1, xmlslz_2, xmlslz_3, TW99, TW99zenodo}   &1.45 &1.323 & 2.63 &3.00 &1.85\\
    QHC21AT       & 1.50   & 1.359 & 2.69  & 3.00 & 1.84 & XMLSLZ(TW99)   &1.50  &1.363 & 2.71 &3.19 &1.85\\
    QHC21AT       & 1.55   & 1.395 & 2.77  & 3.16 & 1.97 & XMLSLZ(TW99)   &1.55 &1.412 & 2.79 &-- &1.84\\
    \hline
    QHC21BT\cite{Kojo2022}       & 1.45   & 1.318 & 2.62  & 2.84 & 1.87 & XMLSLZ(MTVTC)\cite{mtvtc, mtvtczenodo}  &1.45 &1.331 & 2.64 &2.87 &1.68\\
    QHC21BT       & 1.50   & 1.358 & 2.70  & 2.94 & 1.83 & XMLSLZ(MTVTC)  &1.50  &1.377 & 2.73 &3.03 &1.61\\
    QHC21BT       & 1.55   & 1.399 & 2.78  & 3.03 & 1.84 & XMLSLZ(MTVTC)  &1.55 &1.414 & 2.81 &-- &1.59\\
    \hline
    QHC21CT\cite{Kojo2022}       & 1.45   & 1.326 & 2.61  & 2.87 & 1.91 & XMLSLZ(PKDD)\cite{pkdd, pkddzenodo}   &1.45 &1.338 & 2.65 & &\\
    QHC21CT       & 1.50   & 1.368 & 2.69  & 2.97 & 1.91 & XMLSLZ(PKDD)   &1.50  &1.383 & 2.74 &2.58 &1.61\\
    QHC21CT       & 1.55   & 1.410 & 2.77  & 3.07 & 1.98 & XMLSLZ(PKDD)   &1.55 &1.427 & 2.82 &2.61 &1.52\\
    \hline
    QHC21DT\cite{Kojo2022}       & 1.45   & 1.317 & 2.61  & 2.87 & 1.95 & XMLSLZ(GM1)\cite{gm1, gm1zenodo}    &1.45 &1.336 & 2.65 &2.45 &1.45\\
    QHC21DT       & 1.50   & 1.365 & 2.69  & 2.94 & 1.90 & XMLSLZ(GM1)    &1.50  &1.384 & 2.73 &2.49 &1.56\\
    QHC21DT       & 1.55   & 1.412 & 2.77  & 3.00 & 1.86 & XMLSLZ(GM1)    &1.55 &1.420 & 2.81 &2.58 &1.52\\
    \hline
    QHC19BT\cite{Baym2019}       & 1.45   & 1.319 & 2.61  & 3.13 & 1.93 & XMLSLZ(DD-LZ1)\cite{ddlz1, ddlz1zenodo} &1.45 &1.341 & 2.65 &2.39 &1.55\\
    QHC19BT       & 1.50   & 1.357 & 2.69  & 3.32 & 1.93 & XMLSLZ(DD-LZ1) &1.50  &1.389 & 2.73 &2.45 &1.67\\
    QHC19BT       & 1.55   & 1.394 & 2.77  & --   & 1.92 & XMLSLZ(DD-LZ1) &1.55 &1.420 & 2.82 &2.52 &1.57\\
    \hline
    QHC19CT\cite{Baym2019}       & 1.45   & 1.323 & 2.61  & 3.03 & 1.91 & XMLSLZ(TM1)\cite{tm1, tm1zenodo}    &1.45 &1.341 & 2.66 &2.45 &1.51\\
    QHC19CT       & 1.50   & 1.355 & 2.69  & 3.13 & 1.85 & XMLSLZ(TM1)    &1.50  &1.383 & 2.75 &2.52 &1.42\\
    QHC19CT       & 1.55   & 1.403 & 2.77  & 3.26 & 1.93 & XMLSLZ(TM1)    &1.55 &1.424 & 2.83 &2.52 &1.59\\
    \hline
    QHC19DT\cite{Baym2019}       & 1.45   & 1.323 & 2.61  & 3.03 & 1.91 & XMLSLZ(PK1)\cite{pkdd, pk1zenodo}    &1.45 &1.345 & 2.67 &2.39 &1.34\\ 
    QHC19DT       & 1.50   & 1.355 & 2.69  & 3.10 & 1.88 & XMLSLZ(PK1)    &1.50  &1.386 & 2.75 &2.45 &1.46\\
    QHC19DT       & 1.55   & 1.403 & 2.77  & 3.16 & 1.88 & XMLSLZ(PK1)    &1.55 &1.427 & 2.83 &2.45 &1.42\\
    \hline
                  &        &       &       &      &      & XMLSLZ(NL3)\cite{nl3, nl3zenodo}    &1.45 &1.343 & 2.66 & &\\
                  &        &       &       &      &      & XMLSLZ(NL3)    &1.50  &1.387 & 2.75 &2.07 &1.33\\
                  &        &       &       &      &      & XMLSLZ(NL3)    &1.55 &1.431 & 2.83 &2.10 &1.44\\
    % ... other rows ...
    
    % ... other rows ...
  \end{tabular}
  \end{ruledtabular}
\end{table*}

%=====================================================================

%*********************************************************************
%******************** THE MODEL *************************************
%***************                                     *****************
\section{The Model}
\label{themodel}

%*********************************************************************
%******************** Equation of State *****************************
%***************                                     *****************
\subsection{Equations of State}
\label{themodel_eos}

A relativistic perfect fluid requires an EoS of the form
$P=P(\rho,\epsilon)$ to close the hydrodynamic equations and make the otherwise underdetermined system of conservation equations solvable, where $P$ is the pressure, $\rho$ is the rest-mass density and $\epsilon$ is the specific internal energy. Rather than employing a tabulated finite-temperature EoS, we adopt the commonly used decomposition into a cold component and a thermal contribution,
\begin{equation}
P(\rho,\epsilon)
=
P_{\rm cold}(\rho)
+
P_{\rm th}(\rho,\epsilon),
\end{equation}
where $P_{\rm cold}$ describes the zero-temperature EoS in beta equilibrium, while
\begin{equation}
P_{\rm th}
=
(\Gamma_{\rm th}-1)\rho(\epsilon-\epsilon_{\rm cold})
\end{equation}
approximates the thermal pressure generated by shock heating. $\Gamma_{\rm th}=1.8$ was applied in our simulations. We adopt a seven-segment piecewise-polytropic fit \cite{Read2009, DePietri_2016} to represent $P_{\rm cold}$.

The present work considers two classes of realistic EoSs: 1) hybrid EoSs that incorporate a crossover PT to quark matter; and 2) purely hadronic EoSs. For the hybrid EoSs, we adopt the QHC models Ref.~\cite{Baym2019, Kojo2022}. QHC models apply a thermodynamically consistent polynomial interpolation that smoothly connects the low-density hadronic EoS (Togashi or $\chi$EFT models) to the high-density quark matter EoS (NJL model). The interpolation is constructed in terms of the pressure $P$ as a function of the baryon chemical potential $\mu_{\rm B}$. It has been shown \cite{Gangopadhyay20} that a simple polynomial interpolation is as good or better than an infinite series Pade' interpolation.

The quark sector in the QHC model is primarily characterized by two parameters: 1) the universal vector coupling $g_V$, which describes repulsive quark interactions; and 2) the diquark pairing strength $H$. These parameters are chosen to satisfy the requirements of thermodynamic stability, causality, and observational constraints from NSs. The effects of the QHC19 on BNS mergers were discussed in our previous work \cite{Atul2022}. Here, we extend that study to the QHC21 EoS, which is generally stiffer and features a lower upper crossover density of $\sim 3.5\,n_{\rm sat}$, compared to $\sim 5\,n_{\rm sat}$ for QHC19. Consequently, the QHC21 EoSs predict larger radii for massive neutron stars ($\sim 2\,M_{\odot}$). This is in better agreement with the NICER observations.

A summary of the EoSs employed in this work is given in Table.~\ref{tab:tab_mass}. This table identifies the associated hadronic portion and various properties of the isolated stars and characteristic frequencies $f_{\rm peak}$ and $f_{\rm max}$.
Within the QHC21 EoSs, the QHC21T and QHC21$\chi$ differ in their low-density hadronic sectors. QHC21T employs the Togashi EoS. which is based on a variational many-body calculation, while QHC21$\chi$ adopts a microscopic $\chi$EFT EoS. Because the $\chi$EFT EoS is stiffer than the Togashi EoS, QHC21$\chi$ generally predicts slightly larger NS radii than QHC21T. The isolated neutron star mass-radius relations for the adopted EoSs are summarized on the right panel in Figure \ref{fig:mr}. The left panel shows the squared sound speed as a function of density. Note that the sound speed for the QHC EoSs have a pronounced peak in the cross-over density range from 2 to 5 $n_{\rm sat}$. The mass-radius relations in the right panel show that the equation of state span the allowed range of NS radii while maintaining a maximum neutron star mass $\ge 2.08 \pm 0.07$ M$_\odot$ as required \cite{Fonseca2021}. 

The representative purely hadronic EoSs considered here span a broad gamut of stiffness. These include the microscopic BL \cite{Bombaci2018, Carreau2019, BLzenodo} and APR models \cite{Akmal1998, Haensel2007, Davis2024, APRzenodo}, the nonrelativistic density functional SLy model \cite{Douchin2001}, the relativistic PCGS(PCSB2) model \cite{Hempel2010, Pradhan2022, PCGSzenodo} and relativistic XMLSLZ models \cite{xmlslz_1, xmlslz_2, xmlslz_3}, which is based on several covariant density functionals, including TW99 \cite{TW99, TW99zenodo}, MVTVC \cite{mtvtc, mtvtczenodo}, PKDD \cite{pkdd, pkddzenodo}, GM1 \cite{gm1, gm1zenodo}, DD-LZ1 \cite{ddlz1, ddlz1zenodo}, TM1 \cite{tm1, tm1zenodo}, PK1 \cite{pkdd, pk1zenodo}, and NL3 \cite{nl3, nl3zenodo}. These are listed in right half of Table~\ref{tab:tab_mass}.

%One distinction worth noting is that Fig.~\ref{fig:mr} shows the EoSs before the piecewise-polytropic approximation is applied, whereas all subsequent figures and results are based on calculations using their piecewise-polytropic representations. In particular, for the XMLSLZ(DD-LZ1) model, the piecewise-polytropic approximation does not reproduce the characteristic backward-bending, S-shaped behavior of the original mass--radius relation particularly well.

%*********************************************************************
%****** Numerical Method: GRHD Simulations ***************************
%***************                                     *****************
\subsection{Numerical Methods: GRHD Simulations}
\label{themodel_hydrodynamics}

Both the initial data and the hydrodynamic evolution are based on open-source numerical relativity softwares. The initial data for the NS binaries are generated using the LORENE software \cite{Lorene1PRD2001Eric, Lorene2Grandclément2009} under the assumption of irrotational binaries. We consider equal-mass binaries with individual baryonic masses of $1.45$, $1.50$, and $1.55~M_{\odot}$, corresponding to gravitational masses in the range $M_{\rm grav}\sim 1.30$--$1.40~M_{\odot}$. The initial coordinate separation between the stellar centers is set to $45~{\rm km}$. The baryonic masses $M_{\rm baryon}$, along with the corresponding gravitational masses $M_{\rm grav}$ and ADM masses $M_{\rm ADM}$, are summarized in Table~\ref{tab:tab_mass}.

The merger simulations are performed using the Einstein Toolkit numerical relativity software \cite{ET2025}. We adopt Cactus units (CU) with $c=G=M_{\odot}=1$. Adaptive mesh refinement is implemented using the Carpet thorn \cite{Carpet1Schnetter_2006, Carpet2Erik_2004}, with six refinement levels and a finest grid spacing of $0.3125$ CU ($461~{\rm m}$). General relativistic hydrodynamics is evolved using the GRHydro thorn \cite{GRHydro1PRD2005Baiotti, GRHydro2PRD2005Hawke, GRHydro3Mosta2013}, based on the Valencia formulation \cite{Valencia1Font2008, Valencia2Banyuls_1997}. The spacetime metric is evolved using the McLachlan thorn within the BSSN-NOK formalism \cite{GR1, GR2ShibataPRD1995, GR3BaumgartePRD1998, GR4AlcubierrePRD2000, GR5AlcubierrePRD2003}. Time integration is performed using a fourth-order Runge--Kutta method. The HLLE approximate Riemann solver \cite{HLLE} is employed to compute the numerical fluxes, while a fifth-order WENO-Z reconstruction \cite{WENOZ} is used to reconstruct the hydrodynamic variables at cell interfaces.

\subsection{Post-merger GW Analysis}
\label{themodel_gw}
The Newman--Penrose formalism \cite{Newman1962} is used here to extract the GW signal. The Weyl curvature scalar $\Psi_4$ describes the outgoing gravitational radiation field in the asymptotic limit. In our simulation analysis, the GW signal was extracted at a radius of 700 CU (approximately $1000~ \mathrm{km}$), which is considered to be sufficiently far from the source. In this limit, $\Psi_4$ can be directly related to the metric perturbation in the transverse-traceless (TT) gauge and decomposed into spin-weighted spherical harmonics \cite{Bishop2016},
\begin{equation}
\ddot{h}_{+}(t)-i\ddot{h}_{\times}(t)
=\psi_4(t)
=\sum_{\ell=2}^{\infty}\sum_{m=-\ell}^{\ell}
\psi_4^{\ell m}(t)\,
{}_{-2}Y_{\ell m}(\theta,\varphi),
\end{equation}
% of spin-weighted spherical harmonics oscillators.
where $h_{+}$ and $h_{\times}$ are the two polarizations of the GW strain in the face-on direction, $(\theta,\varphi)=(0,0)$, and ${}_{-2}Y_{\ell m}(\theta,\varphi)$ are $s=-2$ spin-weighted spherical harmonics. In the following figures, all GW strain amplitudes are rescaled to a source distance of 50 Mpc, and the waveforms are shifted such that $t=0$ corresponds to the merger time, defined as the instant when the strain amplitude reaches its first maximum, where the amplitude is defined as
\begin{equation}
|h|=\sqrt{h_{+}^{2}+h_{\times}^{2}}.
\end{equation}

%%%%%%%%%%%%%%%%%%%%%%%%%%%%%%%%%%%%%%%%%%%%%%%%%%%%%%%%%%%%%%%%%%%
\begin{figure}[htbp]
\centering
\includegraphics[width=\linewidth]{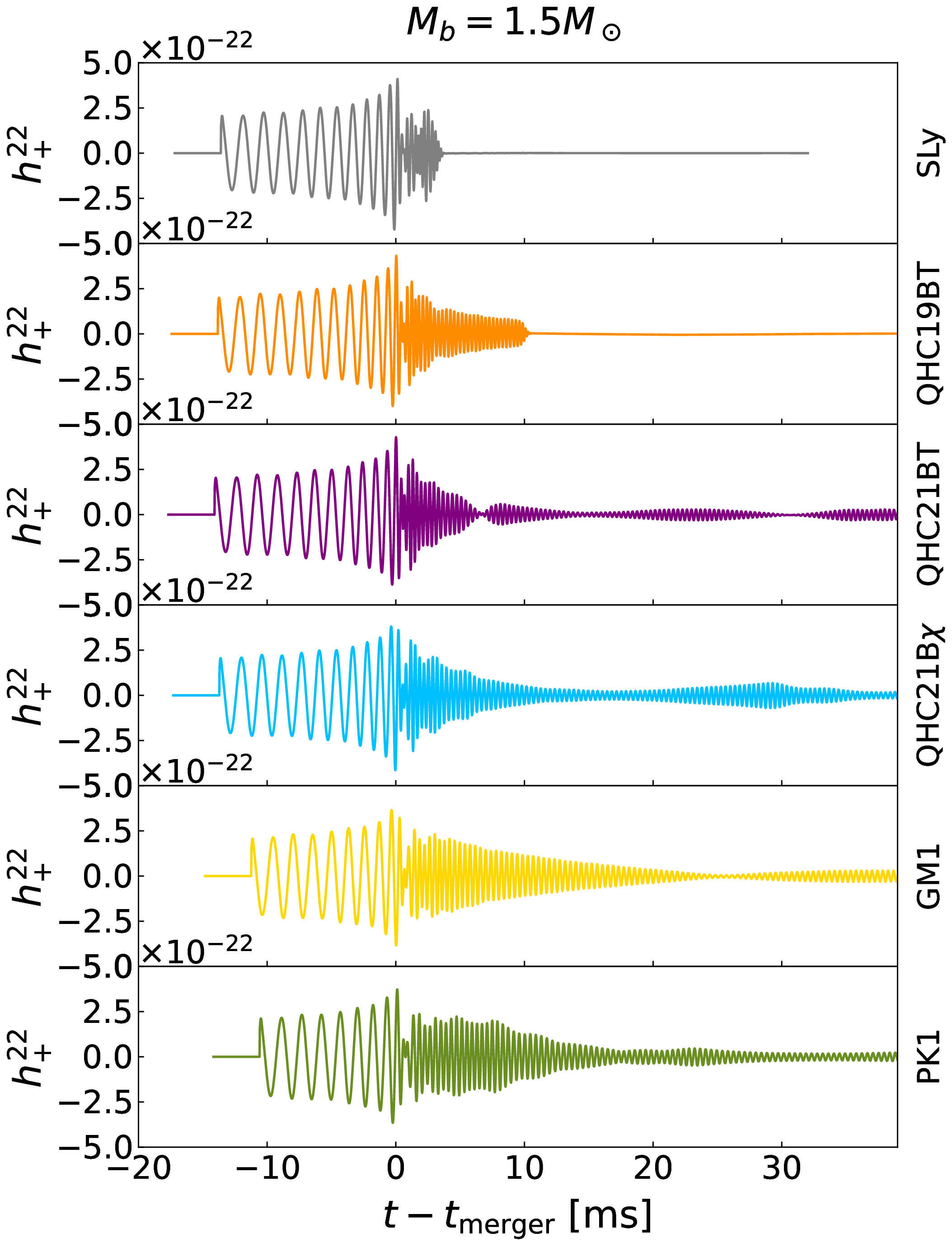}
\caption{\label{fig:strain_several}Comparison of the GW waveforms for six equal-mass binary neutron star mergers with a baryonic mass of $M_{\rm baryon}=1.5\,M_{\odot}$ per star. The panels, arranged from top to bottom in order of increasing EoS stiffness, correspond to the SLy4, QHC19BT, QHC21BT, QHC21B$\chi$, GM1, and PK1 EoSs. The SLy, GM1, and PK1 models are purely hadronic EoSs, whereas the remaining models are crossover hybrid EoSs. The color scheme is consistent with that in Fig.~\ref{fig:mr} and is used to represent different EoSs. }
\end{figure}
%%%%%%%%%%%%%%%%%%%%%%%%%%%%%%%%%%%%%%%%%%%%%%%%%%%%%%%%%%%%%%%%%%%
%%%%%%%%%%%%%%%%%%%%%%%%%%%%%%%%%%%%%%%%%%%%%%%%%%%%%%%%%%%%%%%%%%%
%PSD
\begin{figure*}[htbp]
\centering
\includegraphics[width=\textwidth]{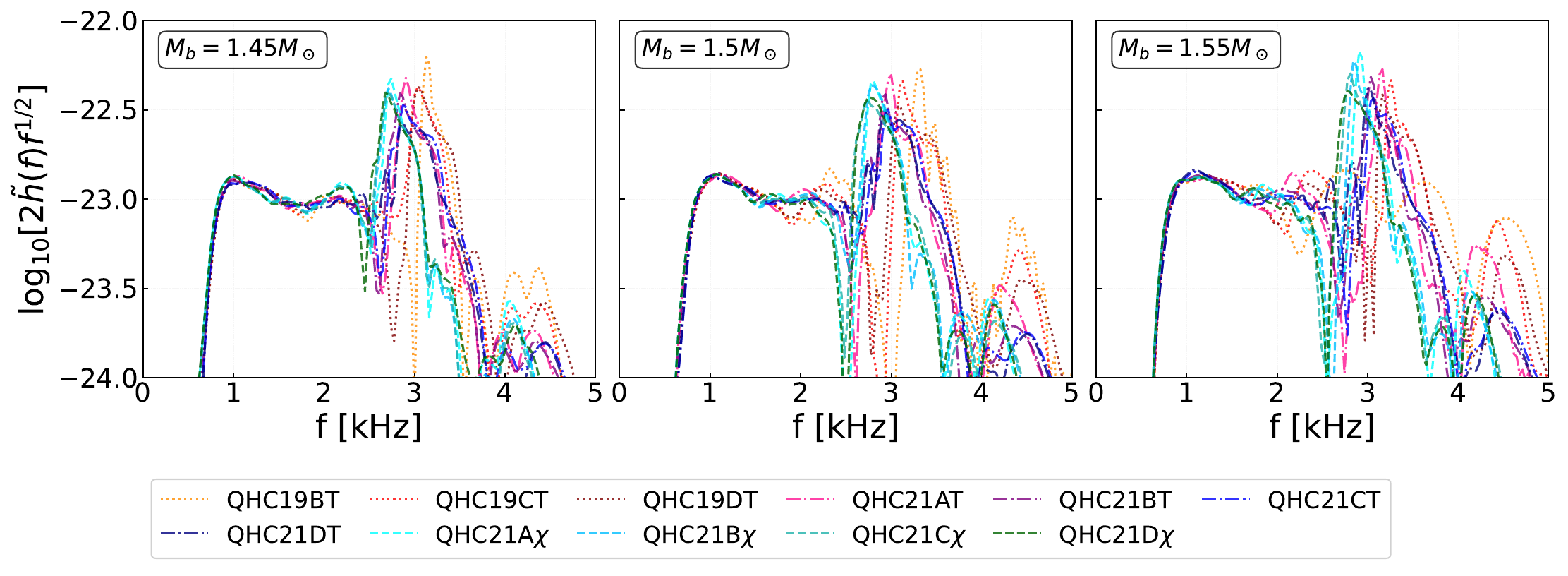}
\vspace{0.3cm} % adjust spacing as needed
\includegraphics[width=\textwidth]{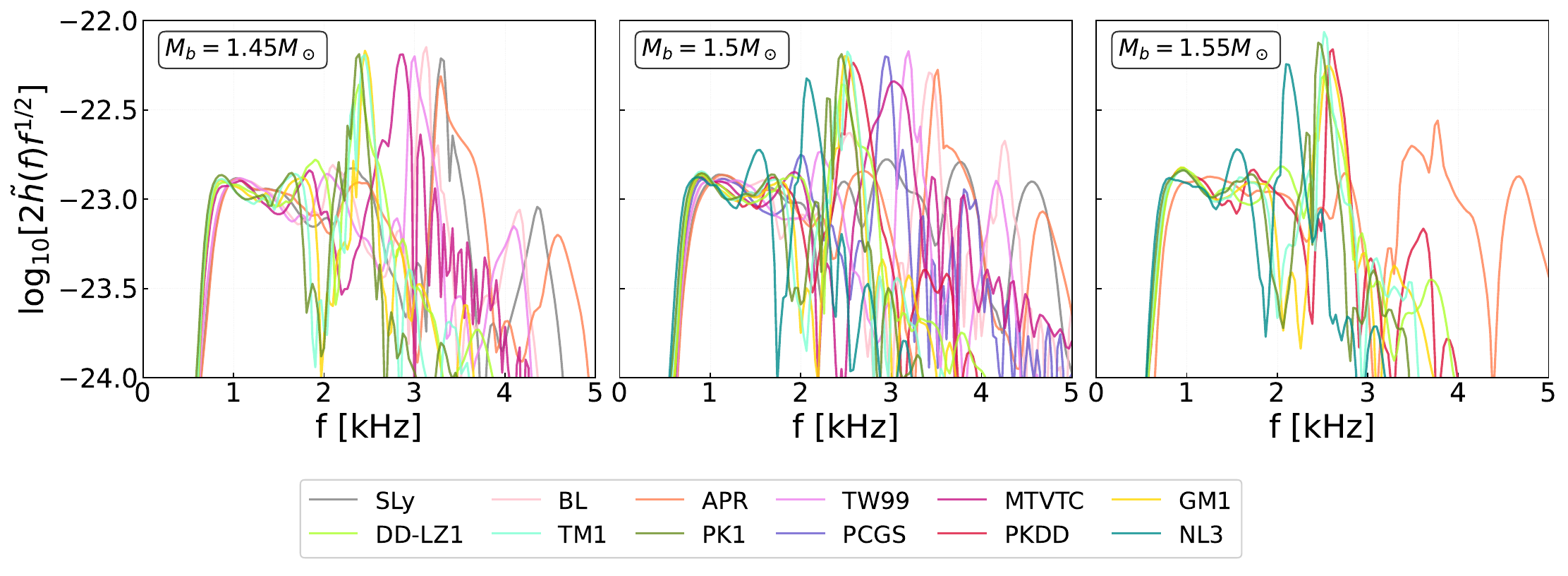}
\caption{Power spectral density for equal-mass binaries with baryonic masses of $M_{\rm baryon}=1.45$, $1.50$, and $1.55\,M_{\odot}$ per star. The top row presents the crossover EoSs, while the bottom row shows the purely hadronic EoSs. The dominant post-merger GW frequency, $f_{\rm peak}$, is identified as the principal peak in the PSD at frequencies of approximately $2$--$3~\mathrm{kHz}$. Only simulations that produce long-lived HMNS are shown, as prompt or early collapse to a BH does not yield a well-defined $f_{\rm peak}$. }
% This is only l=m=2 mode. But since l=m=2 is dominant there is almost no difference between l=m=2 mode and total h. (I've compared. We are using blackman window. Window is [-11,20ms]. Compared with Tukey window with alpha=0.25. Tukey has more zigzag.
\label{fig:psd}
\end{figure*}

\begin{figure*}[ht]
\centering
\includegraphics[width=0.95\linewidth]{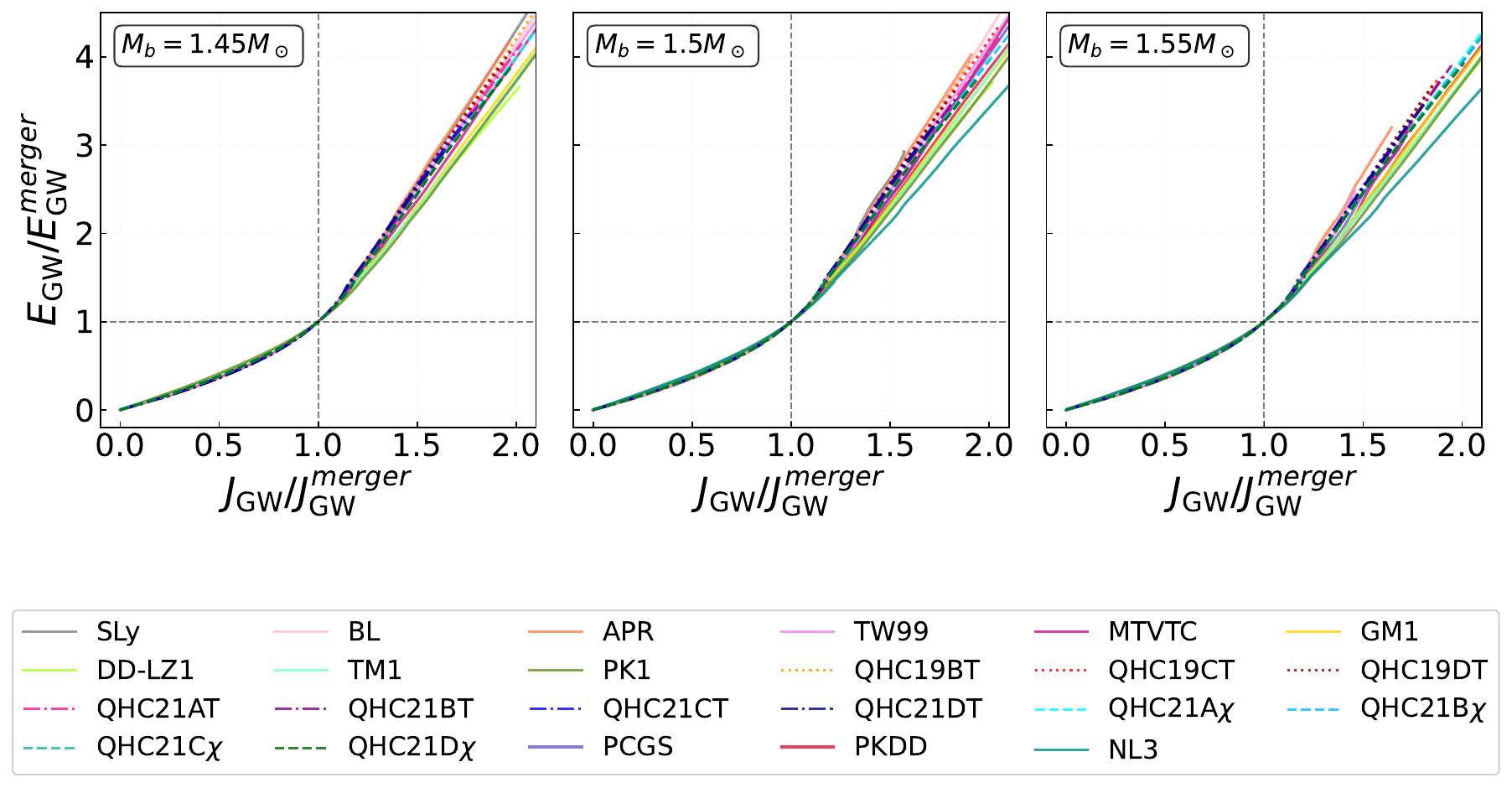}
\caption{\label{fig:EvsJ}Accumulated radiated GW energy $E_{\rm GW}$ as a function of accumulated radiated GW angular momentum $J_{\rm GW}$ for equal-mass binaries with baryonic masses $M_{\rm baryon}=1.45$, $1.50$, and $1.55,M_{\odot}$ per star. Both quantities are normalized by their respective values accumulated from the start of the simulation to merger.
}
\end{figure*}
%%%%%%%%%%%%%%%%%%%%%%%%%%%%%%%%%%%%%%%%%%%%%%%%%%%%%%%%%%%%%%%%%%%

The energy and angular momentum carried by the gravitational waves can also be expressed in terms of $\Psi_4$ \cite{Bishop2016}, where $\bar{\Psi}_4$ denotes the complex conjugate of $\Psi_4$:
\begin{equation}
\frac{dE}{dt}
=
\frac{r^2}{16\pi}
\sum_{\ell=2}^{\infty}
\sum_{m=-\ell}^{\ell}
\left|
\int_{-\infty}^{t}
dt'\,
\psi_4^{\ell m}
\right|^2,
\end{equation}
and
\begin{equation}
\begin{split}
\frac{dJ_z}{dt}
&=
\frac{r^2}{16\pi}
\operatorname{Im}
\Biggl\{
\sum_{\ell=2}^{\infty}
\sum_{m=-\ell}^{\ell}
m
\left(
\int_{-\infty}^{t}
dt'\,\bar{\psi}_4^{\ell m}
\right)
\\
&\qquad\qquad\times
\int_{-\infty}^{t}
dt'
\int_{-\infty}^{t'}
dt''\,\psi_4^{\ell m}
\Biggr\}.
\end{split}
\end{equation}

The instantaneous GW frequency is defined as
\begin{equation}
f_{\rm GW}
=
\frac{1}{2\pi}
\frac{d\phi}{dt},
\qquad
\phi
=
\arctan\!\left(
\frac{h_{\times}}
{h_{+}}
\right).
\end{equation}
The frequency at maximum amplitude is then defined as
\begin{equation}
f_{\rm max}=f_{\rm GW}(t=0).
\end{equation}

To define the signal-to-noise ratio (SNR), we first introduce the Fourier-domain GW amplitude,
\begin{equation}
\tilde{h}(f)
=
\sqrt{
\frac{
\left|\tilde{h}_{+}(f)\right|^{2}
+
\left|\tilde{h}_{\times}(f)\right|^{2}
}{2}
},
\end{equation}
where
\begin{equation}
\tilde{h}_{+,\times}(f)
\equiv
\begin{cases}
\displaystyle
\int_{-\infty}^{\infty}
h_{+,\times}(t)\,
e^{-i2\pi f t}\,dt,
& f\ge0,\\[1ex]
0,
& f<0.
\end{cases}
\end{equation}

Although $|\tilde{h}(f)|$ provides a simple measure of the GW signal in the frequency domain, the signal amplitude can be orders of magnitude below the detector noise level. We therefore introduce the characteristic strain $h_c(f)$ and the noise amplitude $h_n(f)$, both of which are dimensionless:
\begin{equation}
\begin{aligned}
\left[h_c(f)\right]^2
&=
4f^2\left|\tilde{h}(f)\right|^2
=
fS_h(f),\\
\left[h_n(f)\right]^2
&=
fS_n(f).
\end{aligned}
\end{equation}
Here, $S_n(f)$ is the one-sided noise power spectral density (PSD).

The squared SNR is then given by \cite{Moore2015}
\begin{equation}
\begin{aligned}
\rm SNR^2
&=
\int_{0}^{\infty}
d(\log f)\,
\left[
\frac{h_c(f)}
{h_n(f)}
\right]^2
=
\int_{0}^{\infty}
\frac{df}{f}
\frac{S_h(f)}
{S_n(f)}
\\
&=
\int_{0}^{\infty}
\frac{df}{f}
\frac{
\left(2f^{1/2}|\tilde{h}(f)|\right)^2
}
{S_n(f)}.
\end{aligned}
\end{equation}
The square root of the PSD $\sqrt{S_h(f)}$ is shown in Fig.~\ref{fig:psd} and will be discussed in the next section.

%%%%%%%%%%%%%%%%%%%
%%%%%%%%%%%%%%%%%%
\section{Results}
\label{results}
% waveform and HMNS discussion
Figure \ref{fig:strain_several} shows the gravitational-wave waveforms $h^{2,2}_+$, for equal baryon mass $1.5$--$1.5\,M_{\odot}$ systems using several representative EoSs that span the range from soft to stiff behavior for both hadronic and hybrid models. The selected EoSs include three hadronic EoSs: the soft SLy and the relatively stiff GM1 and PK1; and three hybrid EoSs: the relatively soft QHC19BT together with the relatively stiff QHC21BT and QHC21B$\chi$.

The stiffness of an EoS is also reflected in the lifetime of the post-merger HMNS remnant. The softer EoS models like SLy and QHC19BT produce short-lived HMNSs that collapse into BHs shortly ($\lesssim 10$ ms) after merger. In both cases, the pressure support is insufficient to counteract gravitational collapse. 

In contrast, the hybrid models QHC21BT and QHC21B$\chi$, which undergo an earlier crossover to quark matter, remain sufficiently stiff at high densities to provide additional pressure support. As a result, they produce significantly longer-lived HMNS remnants. This is consistent with the previous conclusion in Ref.~\cite{Atul2022} that a crossover to quark matter can prolong the HMNS lifetime. However, it should be kept in mind that a sufficiently stiff hadronic EoS can also produce a similar effect \cite{Takami2015}.

\subsection{Power Spectral Density}
\label{results_psd}

% PSD discussion
EoSs that produce long-lived post-merger remnants give rise to a well-defined dominant GW frequency, $f_{\rm peak}$ \cite{Stergioulas2011}, in the square root of the PSD, as shown in Fig.~\ref{fig:psd}. The quantity $f_{\rm peak}$ has been found to correlate strongly with the properties of the EoS \cite{Bauswein2012, Bauswein2015, Takami2015, Bernuzzi2015, Vretinaris2020, Huang2022, Raithel2022, Breschi2024, Prakash2024, Vretinaris2026}. The top row of Fig.~\ref{fig:psd} presents the crossover EoSs, while the bottom row shows the purely hadronic EoSs. The three columns correspond to binaries of equal-mass with component baryonic masses of 1.45, 1.50, and $1.55\,M_{\odot}$, respectively. The color scheme for each EoS is the same as in the previous figures.

A correlation can be discerned between the stiffness of the EoS and $f_{\rm peak}$. Stiffer EoSs produce lower peak frequencies, whereas softer EoSs shift $f_{\rm peak}$ to higher frequencies. This trend follows from the fact that a stiffer EoS produces a larger remnant radius $R$. A larger radius corresponds to a lower characteristic dynamical 
frequency, $f_{\rm dyn} \sim \sqrt{\frac{GM}{R^{3}}}$ \cite{Nils1998, Misner1973}. Therefore, a stiffer EoS corresponds to a lower $f_{\rm peak}$. Similarly, increasing mass for a given EoS, corresponds to a larger value of $f_{\rm peak}$. This is apparent in Table~\ref{tab:tab_mass}. 

Figure~\ref{fig:EvsJ} shows the accumulated energy radiated in gravitational waves $E_{\rm GW}$, as a function of the accumulated radiated angular momentum $J_{\rm GW}$. A similar approximately linear relation between $E_{\rm GW}$ and $J_{\rm GW}$ has been reported in previous studies \cite{Bernuzzi2016,Bernuzzi2015,Dietrich2017,Ecker2025}. This behavior can be understood from the quadrupole approximation. For a rotating non-axisymmetric system, the GW luminosity scales as $dE_{\rm GW}/dt \sim I^2\Omega^6$, whereas the angular-momentum flux scales as $dJ_{\rm GW}/dt \sim I^2\Omega^5$. Their ratio therefore gives $dE_{\rm GW}/dJ_{\rm GW} \sim \Omega$, where $\Omega$ is the characteristic rotational angular frequency. For a quadrupole $\ell=m=2$ deformation, the GW angular frequency satisfies $\omega_{\rm GW}=2\Omega$, such that $dE_{\rm GW}/dJ_{\rm GW} \sim \Omega = \omega_{\rm GW}/2 = \pi f_{\rm GW}$. This behavior can also be seen in Fig.~1 of Ref.~\cite{Bernuzzi2015} and as discussed in Appendix~A of Ref.~\cite{Takami2015}. Consequently, since stiffer EoSs generally yield lower $f_{\rm peak}$, they are also expected to exhibit smaller values of $dE_{\rm GW}/dJ_{\rm GW}$.

\subsection{Universality Relations}
\label{results_universality}

\begin{figure}[htbp]
\centering
\includegraphics[width=\linewidth]{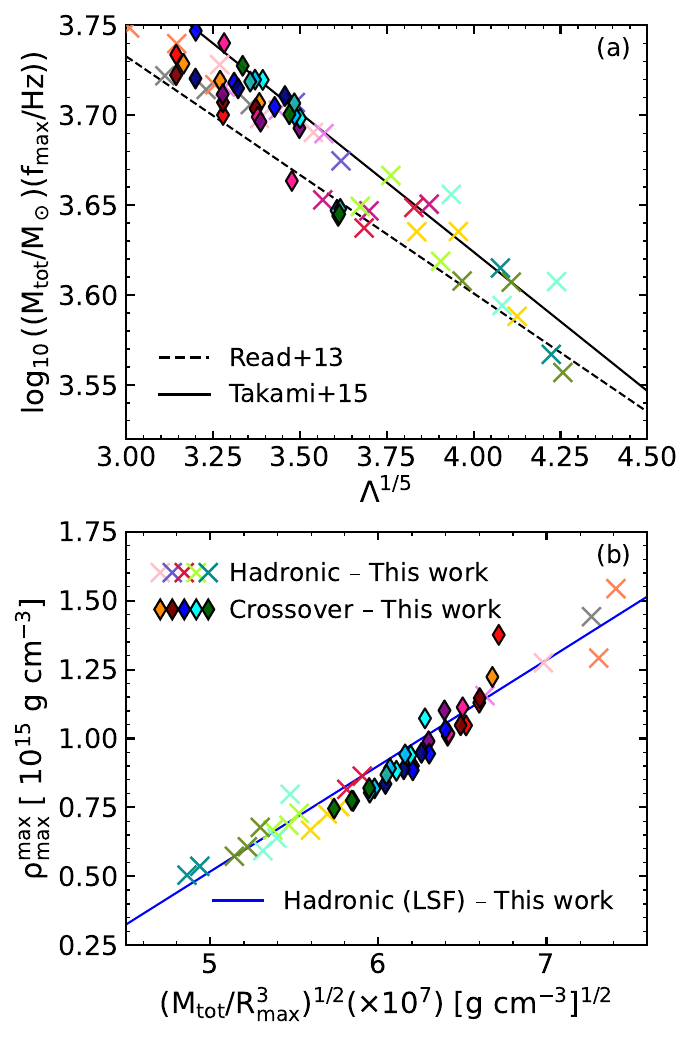}
\caption{\label{fig:uni1}Universality relations between (a) $\log_{10}(M_{\mathrm{tot}}f_{\rm max})$ and $\Lambda^{1/5}$; and (b) $\rho_{\mathrm{max}}^{\mathrm{max}}$ and $(M_{\mathrm{tot}}/R_{\mathrm{max}}^{3})^{1/2}$. The hadronic and crossover PT EoSs considered in this work are compared with the previously proposed universality relations from Read+13 \cite{Read2013} and Takami+15 \cite{Takami2015}.
}
\end{figure}

\begin{figure}[htbp]
\centering
\includegraphics[width=\linewidth]{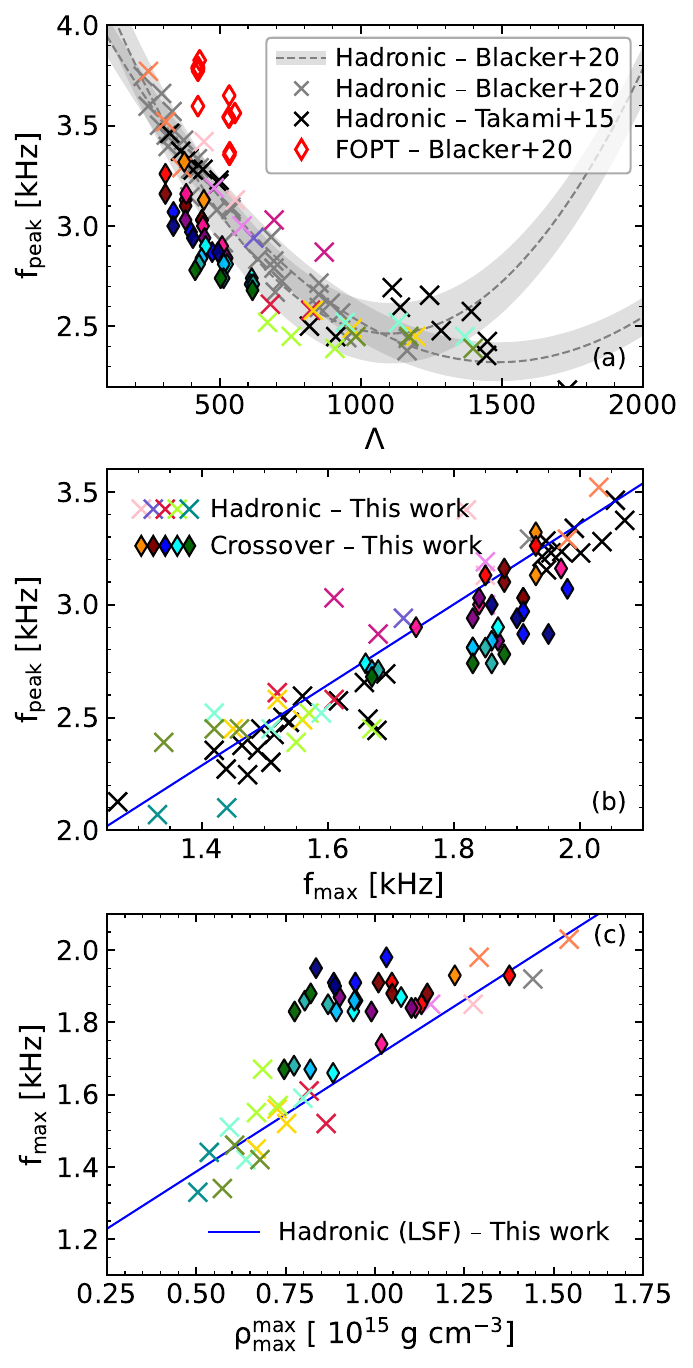}
\caption{\label{fig:uni2}Universality relations between (a) $f_{\mathrm{peak}}$ and $\Lambda$; (b) $f_{\mathrm{peak}}$ and $f_{\mathrm{max}}$; and (c) $f_{\mathrm{max}}$ and $\rho_{\mathrm{max}}^{\mathrm{max}}$. The hadronic EoSs and crossover PT EoSs considered in this work are compared with the previously proposed universality relations and FOPT results from Blacker+20 \cite{Bauswein2020}, as well as the hadronic results from Takami+15 \cite{Takami2015}.
}
\end{figure}

\begin{figure*}[htbp]
\centering
\includegraphics[width=\linewidth]{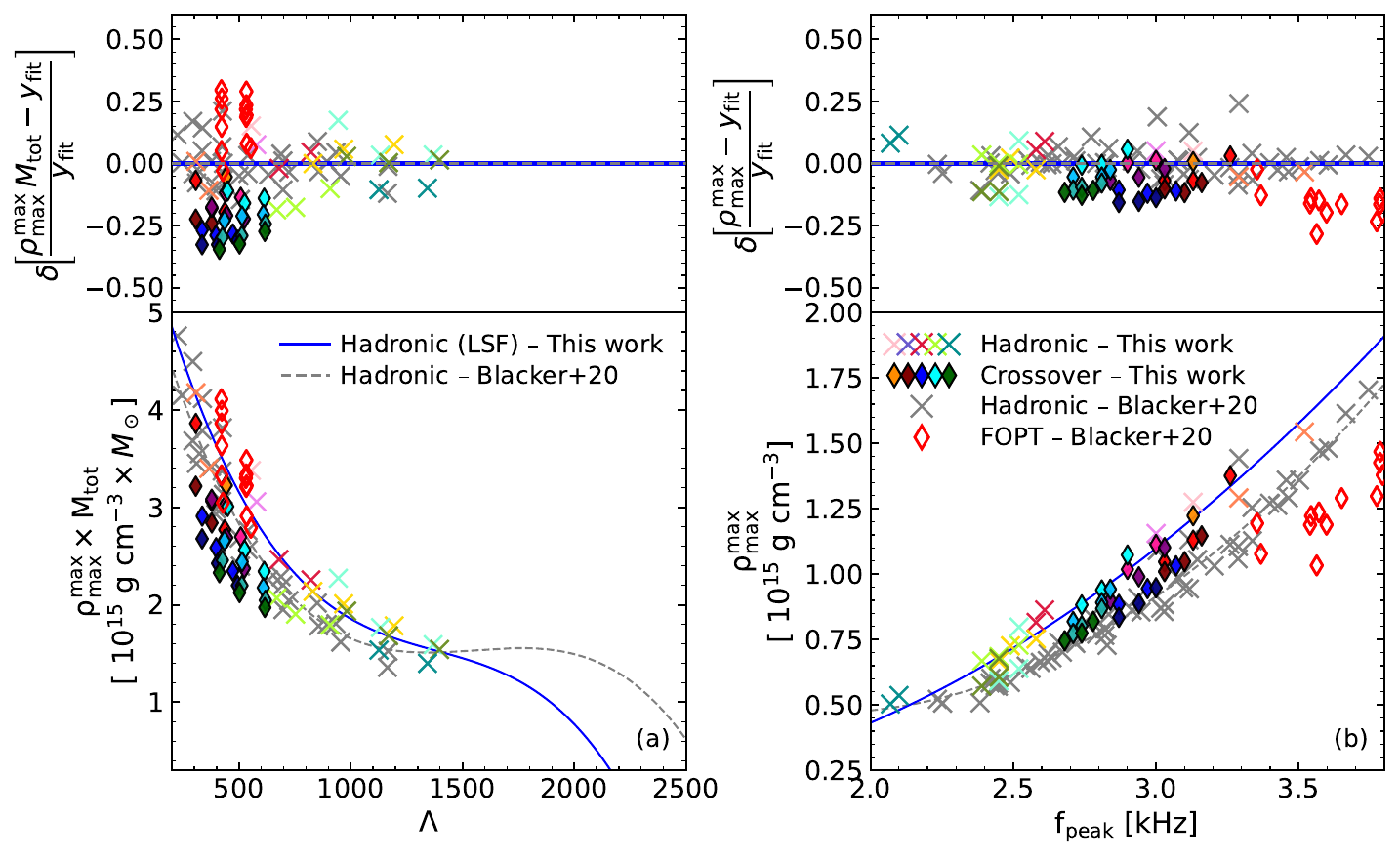}
\caption{\label{fig:uni3}Universality relations between (a) $\rho_{\mathrm{max}}^{\mathrm{max}} M_{\mathrm{tot}}$ and $\Lambda$; and (b) $\rho_{\mathrm{max}}^{\mathrm{max}}$ and $f_{\mathrm{peak}}$. The hadronic and crossover PT EoSs considered in this work are compared with the previously proposed universality relations and FOPT results from Blacker+20 \cite{Bauswein2020}. Here, $y_{\rm fit}$ represents either the least-squares fit from the present work or the hadronic fit from Blacker+20. The top two panels show the deviations of the results for different EoSs from their corresponding fits.
}
\end{figure*}

% universality discussion
Universality relations are a useful way to characterize a broad range of simulation models. In particular, the eventual observation of deviations from such relations in nature may be used as an indicator of new physics, such as a FOPT from hadronic to quark matter \cite{Bauswein2020}. 

Our results on several universality relations are shown in Figures \ref{fig:uni1}, \ref{fig:uni2}, and \ref{fig:uni3},\footnote{Note that, consistent with Table~\ref{tab:tab_mass}, three simulations (PCGS1.45, PKDD1.45, and NL31.45) are not shown due to numerical issues. Seven additional simulations (SLy1.5, SLy1.55, BL1.55, PCGS1.55, TW991.55, MTVTC1.55, and QHC19BT1.55) are excluded from relations involving $f_{\rm peak}$ because the remnants collapse to BHs shortly after merger. Several other simulations (BL1.5, PCGS1.5, TW991.5, MTVTC1.45, MTVTC1.5, QHC19BT1.5, and QHC21AT1.55), which also collapse to BHs within the first 20~ms after merger but persist as HMNSs for an appreciable duration, are retained in the relations involving $f_{\rm peak}$ because their PSDs exhibit a clear contribution from the HMNS phase. These simulations are, however, excluded from relations involving $\rho^{\rm max}_{\rm max}$ because the maximum density increases rapidly to values substantially higher than those attained by stable HMNS remnants.}. We considered more than one universality relation, as they help to clarify the underlying physics of deviations from universality. The hadronic EoSs (colored $\times$ markers) and crossover PT EoSs (colored solid diamonds) calculated in this work are compared with the previously proposed universality relations of Takami+15 \cite{Takami2015} (black $\times$ markers) and Blacker+20 \cite{Bauswein2020} (gray $\times$ markers and bands), together with the FOPT results Blacker+20 ~\cite{Bauswein2020} (red open diamonds).

Figure~\ref{fig:uni1} shows the relations between: (a) $\log_{10}(M_{\mathrm{tot}} f_{\rm max})$ and $\Lambda^{1/5}$; and (b) $\rho_{\mathrm{max}}^{\mathrm{max}}$ and $(M_{\mathrm{tot}}/R_{\mathrm{max}}^{3})^{1/2}$. We note that on these figures the universality relations for the hadronic EoSs also remain valid for the crossover EoSs. Read+13 \cite{Read2013} and Takami+15 \cite{Takami2015} independently obtained the following fitting functions corresponding to the universality relation on panel (a):
\begin{equation}
\log_{10}\left(\frac{M_{\rm tot}}{M_{\odot}}\frac{f_{\rm max}}{\mathrm{Hz}}\right)
= 3.69652 - 0.131743\,\Lambda^{1/5},
\end{equation}
and
\begin{equation}
\log_{10}\left(\frac{M_{\rm tot}}{M_{\odot}}\frac{f_{\rm max}}{\mathrm{Hz}}\right)
= 4.24230 - 0.154600\,\Lambda^{1/5}.
\end{equation}

All hadronic and crossover EoSs considered in the present work fall within the region bounded by these two previously obtained fitting relations in Fig.~\ref{fig:uni1}~(a). Despite differences in the simulation setups among the three studies, this agreement indicates that this universality relation remains robust over the broad range of masses and EoSs considered in this work. We also find that the values of $f_{\rm max}$ and $\Lambda$ for the crossover EoSs are close to those obtained for the soft hadronic EoSs. The explanation for this is that at the time of merger, quark matter has not yet developed sufficiently to significantly alter the bulk properties of the less massive NSs. Consequently, their bulk properties remain similar to those of relatively soft purely hadronic stars.

In panel (b), the least-squares fit (LSF) to the hadronic EoSs yields the linear relation
\begin{equation}
\rho_{\rm max}^{\rm max}
= 0.3835\,\left(\frac{M_{\rm tot}}{R_{\rm max}^{3}}\right)^{1/2} - 1.3998.
\end{equation}
The crossover EoSs also follow this linear relation. While the QHC results are located close to those of the soft hadronic EoSs in Fig.~\ref{fig:uni1}~(a), they lie between the soft and stiff hadronic results in the present relation in Fig.~\ref{fig:uni1}~(b).

In contrast, Figure \ref{fig:uni2} indicates possible deviations from the hadronic universality relations for EoSs with a crossover PT. Panel (a) shows the $f_{\mathrm{peak}}$--$\Lambda$ relation. The results from the present work are compared with those from Blacker+20 \cite{Bauswein2020}, including hadronic EoSs (gray $\times$ markers and bands) and EoSs with a FOPT (red open diamonds), as well as the hadronic results from Takami+15 \cite{Takami2015} (black $\times$ markers). 

The two gray bands represent the hadronic universality relations obtained in Ref.~\cite{Bauswein2020} for equal-mass binaries with component gravitational masses of $1.35\,M_{\odot}$ and $1.40\,M_{\odot}$, respectively. Despite differences in the numerical codes and the adopted thermal parameter $\Gamma_{\mathrm{th}}$ among the different studies, the hadronic simulations consistently fall within the same hadronic universality band. In contrast to the hadronic results, the FOPT EoSs from Ref.~\cite{Bauswein2020} exhibit a clear upward deviation from the hadronic universality band, whereas the crossover EoSs in our simulations show a systematic downward deviation. For both deviations this behavior is consistent with expectations based on the characteristic dynamical frequency that we discussed in Sec.~\ref{results_psd}: softer EoSs (e.g. FOPT EoSs) generally shift $f_{\mathrm{peak}}$ toward higher frequencies, whereas stiffer EoSs (e.g. Crossover EoSs) shift it toward lower frequencies.

Panels (b) and (c), show the $f_{\mathrm{peak}}-f_{\mathrm{max}}$ and $f_{\mathrm{max}}-\rho_{\mathrm{max}}^{\mathrm{max}}$ relations, respectively. These also exhibit clear deviations from the hadronic universality relations for the crossover EoSs. In both cases, the crossover EoSs form distinct clusters away from the least-squares fit to the hadronic EoSs. Such deviations may therefore serve as indicators of a phase transition. The least-squares fitting relations for the hadronic EoSs in Fig.~\ref{fig:uni2}~(b) and (c) are, respectively,
\begin{equation}
f_{\rm peak}
= 1.7858\,f_{\rm max} - 0.2129,
\end{equation}
and
\begin{equation}
f_{\rm max}
= 0.6333\,\rho_{\rm max}^{\rm max} + 1.0705.
\end{equation}

Figure~\ref{fig:uni3} presents $\rho_{\rm max}^{\rm max}$-related universality relations that might serve as indicators for distinguishing FOPT, as suggested by Ref.~\cite{Bauswein2020}. Panel (a) shows the $\rho_{\mathrm{max}}^{\mathrm{max}}M_{\mathrm{tot}}-\Lambda$ relation, while panel (b) shows the $\rho_{\mathrm{max}}^{\mathrm{max}}-f_{\mathrm{peak}}$ relation. In Ref.~\cite{Bauswein2020}, a tight hadronic universality relation was reported for the $\rho_{\mathrm{max}}^{\mathrm{max}}-f_{\mathrm{peak}}$ relation shown in Fig.~\ref{fig:uni3}(b), with the FOPT results exhibiting clear deviations from the universality relation. However, corresponding deviations were not found for the crossover EoSs calculated in the present work.  %models with $M_{\rm grav}=1.35$ and $1.40~M_{\odot}$, whereas the results for $M_{\rm grav}=1.20~M_{\odot}$ shown in Fig.~4(a) of Ref.~\cite{Bauswein2020} are not clearly distinguishable from the corresponding hadronic universality relation.

An explanation for the behavior in Fig.~\ref{fig:uni3}(b) can be explored from the behavior of our crossover EoSs in the other universality relations. The values of $f_{\rm max}$ and $\Lambda$ for the crossover EoSs are found to be close to those of the soft hadronic EoSs, while the values of $f_{\rm peak}$ and $\rho_{\rm max}^{\rm max}$ are both shifted toward lower values due to the onset of quark matter. This explains the crossover outliers seen in Fig.~\ref{fig:uni2}. In those relations, the quantity along one axis is shifted by the crossover PT, while the quantity along the other axis remains close to the soft hadronic data. However, in Fig.~\ref{fig:uni3}(b), when the quantities along both axes are shifted, the effect of a crossover PT is less apparent.

Figure~\ref{fig:uni3}(a), in contrast, considers the relation between $\Lambda$ and $\rho_{\mathrm{max}}^{\mathrm{max}}$. The value of $\Lambda$ remains close to the hadronic values, while $\rho_{\mathrm{max}}^{\mathrm{max}}$ is shifted toward lower values. Therefore, a slight deviation can be seen compared to the fit of the hadronic EoSs. In the present work, however, the fitted hadronic relation differs slightly from that of Ref.~\cite{Bauswein2020}. This discrepancy may arise because the accuracy of $\rho_{\mathrm{max}}^{\mathrm{max}}$ depends sensitively on the simulation code and resolution, particularly since $\rho_{\mathrm{max}}^{\mathrm{max}}$ is measured on the scale of $10^{15}\mathrm{g~cm^{-3}}$. Therefore, we also show the deviations of the results for different EoSs from their corresponding fits, where $y_{\rm fit}$ represents either the fit from Ref.~\cite{Bauswein2020} (gray dashed line) or the fit from the present work (blue solid line). It can be seen that the FOPT results from Ref.~\cite{Bauswein2020} are all above the zero line (except for one point) while the crossover EoSs are all below the zero line. We conclude that the $\rho_{\mathrm{max}}^{\mathrm{max}}M_{\mathrm{tot}}-\Lambda$ relation can also serve as an indicator of a phase transition.

\section{Conclusion}
\label{summary}

We have performed simulations of equal-mass binary neutron stars for a broad range of EoS stiffnesses with and without a crossover phase transition to quark matter. For the hadronic EoSs used as a baseline comparison, different nuclear models were selected, including microscopic calculations, non-relativistic density functional theory, and relativistic density functional theory. The hadronic  models span a broader range from soft to stiff than the crossover EoSs considered in this work. This avoids bias in universality deviations.

The power spectral densities were extracted from the post-merger gravitational-wave emission. A correlation was observed between the stiffness of the EoS and the dominant post-merger GW frequency $f_{\rm peak}$. Stiffer EoSs produce merger remnants with larger radii, resulting in lower characteristic dynamical frequencies. Consequently, they generally yield lower values of $f_{\rm peak}$ and smaller energy emitted in gravitational waves.

A main goal of the present work has been to investigate various universality relations. Among all the relations considered, we found that the $\log_{10}(M_{\mathrm{tot}} f_{\rm max})$-$\Lambda^{1/5}$ relation and the $\rho_{\mathrm{max}}^{\mathrm{max}}$-$(M_{\mathrm{tot}}/R_{\mathrm{max}}^{3})^{1/2}$ relation remain valid for crossover EoSs as well. In contrast, the $f_{\mathrm{peak}}$-$\Lambda$ relation, the $f_{\mathrm{peak}}$-$f_{\mathrm{max}}$ relation, the $f_{\mathrm{max}}$-$\rho_{\mathrm{max}}^{\mathrm{max}}$ relation, and the $\rho_{\mathrm{max}}^{\mathrm{max}} M_{\mathrm{tot}}$-$\Lambda$ relation show clear deviations from the corresponding hadronic universality relations. Therefore they can serve as indicators for distinguishing EoSs with a possible phase transition. In particular, for the $f_{\mathrm{peak}}$-$\Lambda$ relation and $\rho_{\mathrm{max}}^{\mathrm{max}} M_{\mathrm{tot}}$-$\Lambda$ relations, the FOPT and crossover results show deviations in opposite directions from the corresponding hadronic relations.

We also discussed that we did not find deviations for the crossover EoSs in the $\rho_{\mathrm{max}}^{\mathrm{max}}$-$f_{\mathrm{peak}}$ relation. From the relations that remain valid for the crossover EoSs, we found that the values of $f_{\rm max}$ and $\Lambda$ for the crossover EoSs are close to those of the soft hadronic EoSs. This can be understood because these two quantities are primarily sensitive to the relatively low-density regime, where quark matter does not yet play a significant role. In contrast, the values of $f_{\rm peak}$ and $\rho_{\rm max}^{\rm max}$ have both been shifted toward lower values for the crossover EoSs due to the onset of quark matter. 

This difference explains the clear deviations of the crossover EoSs from several universality relations established for the hadronic EoSs. In these relations, the quantity along one axis is shifted by the crossover PT, while the quantity along the other axis remains close to the soft hadronic data. However, when the quantities along both axes are shifted, the shifts can partially compensate for each other, making the deviation from the hadronic universality relation less apparent. This is the reason why our crossover EoSs deviation cannot be seen in the $\rho_{\mathrm{max}}^{\mathrm{max}}$-$f_{\mathrm{peak}}$ relation.

We suggest that future observational determinations of the power spectral density from gravitational wave emission in the 2-3.5 kHz region may show evidence of such deviations from the hadronic universality relations. This could provide evidence for a crossover PT to quark matter and distinguish between a crossover PT and a FOPT.

\section{Acknowledgments}
We thank Hee Il Kim and In-Saeng Suh for helpful discussions during the early phase of this work. Work at the University of Notre Dame is supported by the U.S. Department of Energy under Nuclear Theory Grant DE-FG02-95-ER40934. This research used resources of the National Energy Research Scientific Computing Center, a DOE Office of Science User Facility supported by the Office of Science of the U.S. Department of Energy under Contract No. DE-AC02-05CH11231 using NERSC award NP-ERCAP0036100. A.K. benefited from interactions and workshops supported by the Center for Nuclear astrophysics Across Messengers (CeNAM), which is supported by the U.S.\ DOE, Office of Science, Office of Nuclear Physics, under Award Number DE-SC0023128. Part of the numerical simulations presented in this work were performed using computational resources provided by the University of Notre Dame Center for Research Computing (CRC).
%==============================================================

%*********************************************************************
%******************** ENDING *************************************
%***************                                     *****************
\bibliography{main} 
\end{document}